\documentclass[onecolumn,preprintnumbers,epsfig,amsmath,amssymb]{revtex4}

\usepackage{bm}

\usepackage{epsfig}

\begin{document}

\title{Can We Rigorously Define Phases in a Finite System?}  

\author{K. A. Bugaev$^1$, A. I. Ivanytskyi$^1$, E. G. Nikonov$^2$, A. S. Sorin$^3$ and G. M. Zinovjev$^1$}

\affiliation{$^1$Bogolyubov Institute for Theoretical Physics,
Kiev 03680, Ukraine\\
$^2$Laboratory for Information Technologies, JINR Dubna, 
141980 Dubna, Russia\\
$^3$Bogoliubov  Laboratory of Theoretical Physics, JINR Dubna, 
141980 Dubna, Russia
}

\date{\today}

\begin{abstract}
Here we propose the generalized statistical multifragmentation model  which 
includes the liquid phase pressure of the most general form. This allows us to 
get rid of the absolute  incompressibility of the nuclear liquid. 
Also the present model employes a very general form  the surface tension 
coefficient of nuclear fragments. Such a model is solved analytically for finite volumes
by the Laplace-Fourier transform method to isobaric ensemble. 
A complete analysis of the isobaric partition singularities of this model is also done for finite volumes. 
It is shown that the real part of  any simple pole of the isobaric partition defines the free energy
of the corresponding state, whereas its  imaginary part, depending on the sign,  defines
the inverse decay/formation time of this state. 
The developed  formalism allows us  to exactly define the finite volume analogs
of gaseous, liquid and mixed  phases of the class of similar   models   from the first principles of statistical mechanics and 
demonstrate the pitfalls of  earlier  works. 
The finite
width effects for large nuclear fragments and quark gluon bags  are also  discussed. 

\vspace*{0.2cm}

\noindent
\hspace*{0.0cm}{\bf PACS} numbers: 25.70. Pq, 21.65.+f, 24.10. Pa
\end{abstract}

\maketitle



\section{Introduction}

The necessity to extend the  theory of 1-st order liquid-gas  phase transition  (PT) is determined both  by an academic interest to this problem and by practical purpose to study the phase transformations in the systems that do not have thermodynamic limit. 
Such an extension is  of particular interest for  nuclear physics of intermediate energies 
where the nuclear liquid-gas PT \cite{Bondorf:95,Gross:97,Moretto:97} is studied in the presence of the Coulomb interaction. 
Also recently there appeared a great attention to  PTs  in finite systems because of 
the searches for the new state of matter, the quark gluon plasma, and its (tri)critical  endpoint  in the relativistic collisions 
of heavy ions \cite{Shuryak:09,JINR:08,Bugaev:10}. 

This problem has a long history, but up to now it is not resolved.  One of the first attempts of its resolving 
was made by T. Hill \cite{Hill} whose approach  is based  on the formulation of  thermodynamics of small systems. HillÕs ideas were developed further in \cite{Chomaz:03}, where the authors claimed to establish a one-to-one correspondence between the bimodal structure of the partition of measurable quantity B, known on average, and the properties of the Lee--Yang zeros \cite{Yang-Lee}   of this partition in the complex g-plane. 
However, the analysis of such a definition performed in   \cite{Bugaev:07a} on the basis 
of exactly solvable model for finite volume \cite{Bugaev:05a} shows that the definition of phases suggested in \cite{Chomaz:03} cannot be established experimentally under any circumstances.  Therefore, although  many important  aspects of  the nuclear liquid-gas PT related to finite volumes of studied nuclear systems are well understood 
\cite{Bondorf:95,Moretto:97,Nucleation} the systematic and rigorous extension  of the PT theory to finite systems is just at the very initial stage in spite of  too optimistic  beliefs of Ref. \cite{Chomaz:04}.

Therefore, here we would like to discuss  a powerful mathematical method
invented recently \cite{Bugaev:05a}, the Laplace-Fourier transform (LFT), that 
allows one  not only  to solve  analytically several statistical models for finite volumes or surfaces, but also  to establish a common framework to study 
the deconfinement PT and nuclear liquid-gas PT in finite systems. 
LFT was successfully  applied to the simplified version of the statistical multifragmentation model (SMM) \cite{Gupta:98,Gupta:99} in finite volumes \cite{Bugaev:05a}, to the analysis of  surface partition and surface entropy  of large but finite physical  clusters for  
a variety of statistical ensembles  \cite{Bugaev:04b} and to  exact solution 
of the gas of quark gluon bags \cite{GPZ:81} for  finite volumes \cite{Bugaev:07a}. 
Furthermore, using the theorems proven in \cite{Bugaev:07a,Bugaev:DoS} it is possible 
to  straightforwardly apply  the exact representation of the finite volume  grand canonical partition (GCP) 
of  the gas of quark gluon bags model \cite{Bugaev:07a} 
to other exactly statistical  models of  the deconfinement PT which were solved recently 
in thermodynamic limit \cite{QGBSTM1,QGBSTM2, FWM:08, Reggeons1, Reggeons2}. 
Finally, we would like to stress that our approach to study PTs in finite systems is not restricted to the models without long-range Coulomb-like interaction.  Our strategy is 
as follows: to firmly define the phases in finite systems without the long-range  interaction,
and then to extend this approach to the systems with the Coulomb-like interaction. 
This work is mainly  devoted to the first of these tasks  for which the finite volume solution of the simplified  version of the SMM (and similar models) is  analyzed here. 
In  addition, here  we  generalize 
the simplified version of the SMM \cite{Gupta:98,Gupta:99} in order to repair such  its defects as 
an existence of limiting baryonic density and an absence of (tri)critical endpoint for the values 
of the Fisher exponent $\tau > 2$ \cite{Bugaev:00,Reuter:01}. 

The work is organized as follows. Section II is devoted to the formulation of the generalized 
SMM.  Section III  contains an  introduction  into the LFT technic. The analysis of the singularities of the isobaric partition of the suggested model is given in Section IV. In Sections V and VI we, respectively,  discuss 
the location of the isobaric partition singularities in complex plane for the case without PT and with PT in thermodynamic limit. The finite volume analogs of phases  along with the critique of the Hill's treatment of 1-st order  PT in finite systems are presented in 
Section VI. 
Finally, Section VII contain our concluding remarks.


\section{Generalized SMM}

The system states in the standard  SMM are specified by the multiplicity
sets  $\{n_k\}$
($n_k=0,1,2,...$) of $k$-nucleon fragments.
The partition function of a single fragment with $k$ nucleons is
\cite{Bondorf:95}:
$
V \phi_k (T) = V\left(m T k/2\pi\right)^{3/2}~z_k~
$,
where $k=1,2,...,A$ ($A$ is the total number of nucleons
in the system), $V$ and $T$ are, respectively, the  volume
and the temperature of the system,
$m$ is the nucleon mass.
The first two factors  on the right hand side (r.h.s.) 
of 
the single fragment partition 
originate from the non-relativistic thermal motion
and the last factor,
 $z_k$, represents the intrinsic partition function of the
$k$-nucleon fragment. Therefore, the function $\phi_k (T)$ is a phase space
density of the k-nucleon fragment. 
For \mbox{$k=1$} (nucleon) we take $z_1=4$
(4 internal spin-isospin states)
and for fragments with $k>1$ we use the expression motivated by the
liquid drop model (see details in \mbox{Ref. \cite{Bondorf:95}):}
$
z_k=\exp(-f_k/T),
$ with the  fragment free energy of the simplified SMM
\begin{equation}\label{one}
f_k = - W(T)~k 
+ \sigma (T)~ k^{2/3}+ (\tau + 3/2) T\ln k~,
\end{equation}
with $W(T) = W_{\rm o} + T^2/\epsilon_{\rm o}$.
Here $W_{\rm o}=16$~MeV is the bulk binding energy per nucleon.
$T^2/\epsilon_{\rm o}$ is the contribution of
the excited states taken in the Fermi-gas
approximation ($\epsilon_{\rm o}=16$~MeV). $\sigma (T)$ is the
temperature dependent surface tension parameterized
in the following relation:
$
\sigma (T)=\sigma_{\rm o}
[(T_c^2~-~T^2)/(T_c^2~+~T^2)]^{5/4},
$
with $\sigma_{\rm o}=18$~MeV and $T_c=18$~MeV ($\sigma=0$
at $T \ge T_c$). The last contribution in Eq.~(\ref{one}) involves the Fisher's term with
dimensionless parameter $\tau$. 
The free  energy (\ref{one}) does not contain the symmetry and
Coulomb contributions which are neglected. Hence it is called a simplified version of the SMM
which was suggested and studied numerically in Refs. \cite{Gupta:98,Gupta:99}.
However, its investigation
appears to be of  principal importance
for studies of the liquid-gas phase transition in finite systems.

From an exact analytical solution of the simplified SMM \cite{Bugaev:00} it follows that the baryonic density  has a  maximal  value $\max{\rho} = 1/b = \rho_{\rm o} =0.16 $ fm$^{-3}$  in the limit $\mu \rightarrow \infty$
(here $\rho_{{\rm o}}$ is the normal nuclear density).
This, however, contradicts to the experiments on heavy ion collisions \cite{Pawel:02} in which the nuclei can be compressed to much higher densities. In order to avoid such a behavior we generalize the SMM to the GSMM  which includes the pressure of the liquid phase 
$p_l(T,\mu)$ in a general form. Then the free energy of the fragments with $k>1$ reads as
\begin{equation}\label{EqII}
f_k^G = \mu\, k - p_l(T,\mu) \, b\, k 
+ \sigma (T)~ k^{2/3}+ (\tau + 3/2) T\ln k~,
\end{equation}
where the  pressure of a liquid phase is (at least)  a double differentiable function of its arguments  that  contains the temperature dependent  binding energy $W(T)$. Also  
it is assumed that the  function $p_l(T,\mu)$  reproduces all the typical properties 
of a liquid phase. An extremely important property of the GSMM is  that the liquid phase equation of state  $p_l(T,\mu)$ can be taken from some microscopic models including the
mean-field ones, but the resulting model will be a truly  statistical one since the analytical properties of the isobaric partition  singularities remain unmodified in this case. 

The GSMM nucleons are considered as in the SMM.   Note that the pressure of a liquid phase should approach the asymptotics  $p_l \sim T^2$ for $T\rightarrow \infty$ and 
$p_l \sim \mu^2$ for $\mu \rightarrow \infty$ to respect a causality condition \cite{RelVDW}.
Therefore, the simplest parameterization of the liquid phase pressure which at low densities recovers the usual SMM result and at high densities obeys such asymptotics can be written  as follows 
\begin{equation}\label{EqIII}
 p_l(T,\mu) = \frac{\mu(1 + a\, \mu) + W(T) }{b}\, ~,
\end{equation}
where a positive constant $a > 0$ has to be fixed by the condition that at low densities 
it behaves as $a |\mu| \ll 1$. However, in what follows we study the most general form of the liquid phase pressure. 

In addition to the new parameterization of the free energy of the $k$-nucleon fragment (\ref{EqII}) we  propose to consider a more general parameterization of the surface tension coefficient 
\begin{equation}\label{EqIV}
 \sigma (T) =  \sigma_{\rm o} \left| \frac{T_c - T }{T_c} \right|^\zeta  {\rm sign} ( T_c - T) ~,
\end{equation}
with  $\zeta = const \ge 1$ and $T_c =18$ MeV. In contrast to the Fisher droplet model \cite{Fisher:67} and   the SMM \cite{Bondorf:95}, the GSMM surface tension (\ref{EqIV}) is negative above the critical temperature $T_c$. It is necessary to stress that there is nothing wrong 
or unphysical with the negative values of surface tension coefficient  (\ref{EqIV}), since $ \sigma\, k^\frac{2}{3}$ in (\ref{EqII}) is the surface  free energy of the
fragment  of mean volume $b \,k $ and, hence, as any free energy,  it contains the energy part
$e_{surf}$ and  the entropy part $s_{surf}$ multiplied by temperature $T$ \cite{Fisher:67}. Therefore, at low temperatures the energy part dominates and the  surface  free energy is positive, whereas at high temperatures the number of fragment  configurations with large surface  drastically increases  and it exceeds  the  Boltzmann suppression and, hence,
the surface free energy becomes negative since $s_{surf} > \frac{e_{surf}}{T}$.
Because of this reason   the negative values of the surface tension coefficient  were recently  employed in a variety of exactly solvable statistical models for the deconfinement PT 
\cite{QGBSTM1,QGBSTM2,FWM:08,Aleksey:11}. For the first time  this fact was  derived  within the  exactly solvable models for surface deformations of large physical clusters  \cite{Bugaev:04b}. Very recently two of us 
derived a relation between the surface tension of large quark gluon bags and the string tension  of two  static  color charges measured by the lattice QCD \cite{String:10} from which it was possible to conclude that at high temperatures the surface tension coefficient of quark gluon bags should be negative \cite{String:10,String:11}. 

Furthermore,  a thorough  analysis of the temperature dependence of the surface tension coefficient  in ordinary liquids \cite{KABJGross:09,KABScaling:06} shows not only  that 
the surface tension coefficient approaches zero, but, in contrast to the widely  spread beliefs, for many liquids  the  full $T$ derivative of $\sigma$  does not vanish and remains finite at  $T_{c}$: $\frac{d~ \sigma}{d ~T} < 0$ \cite{KABScaling:06}. Therefore, just the naive extension of these data to the temperatures above  $T_{c}$ would lead to  negative values of  surface tension coefficient at the supercritical temperatures. On the other hand, if one, as usually,   believes that $\sigma \equiv 0$ for $T >T_{c} $, then  it is absolutely unclear what  physical process can lead to  simultaneous existence of  the discontinuity of $\frac{d~ \sigma}{d ~T}$ at 
$T_{c}$ and the  smooth behavior   of the pressure's  first and second derivatives   at the cross-over. Therefore, we conclude that  negative values of the surface tension coefficient   at supercritical temperatures are  also necessary for ordinary liquids although up to now this question has not been  investigated. 

The existence of  negative values of the surface tension coefficient in (\ref{EqIV}) leads  to entirely  new result for the GSMM compared to that one of  the SMM for $\tau > 2$. 
Thus, for $\tau > 2$ the SMM predicts an existence of the 1-st order PT up to infinite values of $T$ \cite{Bugaev:DoS,Bugaev:00, Reuter:01}. Clearly, such a result does not correspond to the experimental findings  and is usually understood as a pitfall of this  model. However, 
the negative values of $\sigma$ in (\ref{EqIV}) lead to a different result in the GSMM. 
Using the technic developed in \cite{QGBSTM1, Aleksey:11} it is easy to show that  in this case  there is a cross-over for $T > T_c$  and, hence, for 
$\tau > 2$ the GSMM has a critical point at $T= T_c$.

\section{The Laplace-Fourier Transformation Technic}

To evaluate the GCP of the GSMM for finite volumes  first we define
the canonical partition function (CPF) of nuclear fragments.  The latter  
has the following form:
\begin{equation} \label{two}
\hspace*{-0.2cm}Z^{id}_A(V,T)=\sum_{\{n_k\}} \biggl[
\prod_{k=1}^{A}\frac{\left[V~\phi_k(T) \right]^{n_k}}{n_k!} \biggr] 
{\textstyle \delta(A-\sum_k kn_k)}\,.
\end{equation}
In Eq. (\ref{two}) the nuclear fragments are treated as point-like objects.
However, these fragments have non-zero proper volumes and
they should not overlap in the coordinate space.
In the excluded volume (Van der Waals) approximation
this is achieved by substituting
the total volume $V$ in Eq. (\ref{two}) by the free (available) volume
$V_f\equiv V-b\sum_k kn_k$, where $b$ is eigen volume of nucleon. 
Therefore, the corrected CPF becomes:
$
Z_A(V,T)=Z^{id}_A(V-bA,T)
$.

The calculation of $Z_A(V,T)$  is difficult due to  the constraint $\sum_k kn_k =A$.
This difficulty can be partly avoided by evaluating
the  GCP function:
\begin{equation}\label{three}
{\cal Z}(V,T,\mu)~\equiv~\sum_{A=0}^{\infty}
\exp\left({\textstyle \frac{\mu A}{T} }\right)
Z_A(V,T)~\Theta (V-bA) ~,
\end{equation}
where $\mu$ denotes a chemical potential.
Nevertheless, the calculation of ${\cal Z}$  is still rather
difficult. The summation over $\{n_k\}$ sets
in $Z_A$ cannot be performed analytically because of
additional $A$-dependence
in the free volume $V_f$ and the restriction
$V_f>0 $.
This problem was resolved   \cite{Bugaev:00} 
by the Laplace transformation method to 
the so-called
isobaric ensemble \cite{GPZ:81}.

To study the PT in finite systems here  we  consider a more strict constraint
$\sum\limits_k^{K(V)} k~n_k =A$ , where the size
of the largest fragment  $K(V) = \alpha V/b$ cannot exceed the total volume of the system 
(the parameter $\alpha \le 1$  is introduced for convenience).
 The case $K(V) = const$ considered in \cite{CGSMM} is also included in our treatment. 
A similar restriction should be also applied to the upper limit of the product in 
all partitions $Z_A^{id} (V,T)$, $Z_A(V,T)$
and ${\cal Z}(V,T,\mu)$ introduced above 
(how to deal with the real values of $K(V)$, see  \cite{Bugaev:05a}). 
Then the  model with such a  constraint, the CGSMM,  cannot be solved by the Laplace 
transform method, because the volume integrals cannot be evaluated due to a complicated 
functional $V$-dependence.  
However, the CGSMM can be solved analytically with the help of  the following identity \cite{Bugaev:05a}
\begin{equation}\label{four}
G (V) = 
\int\limits_{-\infty}^{+\infty} d \xi~ \int\limits_{-\infty}^{+\infty}
  \frac{d \eta}{{2 \pi}} ~ 
{\textstyle e^{ i \eta (V - \xi) } } ~ G(\xi)\,, 
\end{equation}
which is based on the Fourier representation of the Dirac $\delta$-function. 
The representation (\ref{four}) allows us to decouple the additional volume
dependence and reduce it to the exponential one,
which can be dealt by the usual Laplace transformation
in the  following sequence of steps
\begin{eqnarray} \label{five}
\hat{\cal Z}(\lambda,T,\mu) &\equiv &\int_0^{\infty}dV~{\textstyle e^{-\lambda V}}
~{\cal Z}(V,T,\mu) = 
%
%
\int_0^{\infty}\hspace*{-0.2cm}dV^{\prime}
\int\limits_{-\infty}^{+\infty} d \xi~ \int\limits_{-\infty}^{+\infty}
\frac{d \eta}{{2 \pi}} ~ { \textstyle e^{ i \eta (V^\prime - \xi) - \lambda V^{\prime} } } \times 
\nonumber \\
&& 
\sum_{\{n_k\}}\hspace*{-0.cm} \left[\prod_{k=1}^{K( \xi)}~\frac{1}{n_k!}~\left\{V^{\prime}~
{\textstyle \phi_k (T) \,  
e^{\frac{ (\mu  - (\lambda - i\eta) bT )k}{T} }}\right\}^{n_k} \right] \Theta(V^\prime) = \nonumber \\
&&\hspace*{-0.cm}
\int_0^{\infty}\hspace*{-0.2cm}dV^{\prime}
\int\limits_{-\infty}^{+\infty} d \xi~ \int\limits_{-\infty}^{+\infty}
  \frac{d \eta}{{2 \pi}} ~ { \textstyle e^{ i \eta (V^\prime - \xi) - \lambda V^{\prime} 
+ V^\prime {\cal F}(\xi, \lambda - i \eta) } }\,.
%
%
%
\end{eqnarray}
After changing the integration variable $V \rightarrow V^{\prime} = V - b \sum\limits_k^{K( \xi)} k~n_k $,
the constraint of $\Theta$-function has disappeared.
Then all $n_k$ were summed independently leading to the exponential function.
Now the integration over $V^{\prime}$ in Eq.~(\ref{five})
can be straightforwardly done resulting in
\begin{equation}\label{six}
\hspace*{-0.4cm}\hat{\cal Z}(\lambda,T,\mu) = \int\limits_{-\infty}^{+\infty} \hspace*{-0.1cm} d \xi
\int\limits_{-\infty}^{+\infty} \hspace*{-0.1cm}
\frac{d \eta}{{2 \pi}} ~ 
\frac{  \textstyle e^{ - i \eta \xi }  }{{\textstyle \lambda - i\eta ~-~{\cal F}(\xi,\lambda - i\eta)}}~,
\end{equation}

\vspace*{-0.3cm}

\noindent
where the function ${\cal F}(\xi,\tilde\lambda)$ is defined as follows 
\begin{eqnarray}\label{seven}
&&\hspace*{-0.4cm}{\cal F}(\xi,\tilde\lambda) = \sum\limits_{k=1}^{K(\xi) } \phi_k (T) 
~e^{\frac{(\mu  - \tilde\lambda bT)k}{T} } =
\left(\frac{m T }{2 \pi} \right)^{\frac{3}{2} } \hspace*{-0.1cm} \biggl[ z_1
~{\textstyle e^{ \frac{\mu- \tilde\lambda bT}{T} } } + \hspace*{-0.1cm} \sum_{k=2}^{K(\xi) }
k^{-\tau} e^{ \frac{( p_l(T,\mu)- \tilde\lambda T)b k - \sigma k^{2/3}}{T} }  \biggr]\,.\,
\end{eqnarray}
This result generalizes the finite volume solution of the simplified SMM  obtained in \cite{Bugaev:05a,Bugaev:07a}. 

As usual, in order to find the GCP by  the inverse Laplace transformation,
it is necessary to study the structure of singularities of the isobaric partition (\ref{seven}).

\section{Isobaric Partition Singularities}

The isobaric partition (\ref{seven}) of the CGSMM is, of course, more complicated
than its SMM analog found in thermodynamic limit  \cite{Bugaev:00}
because for finite volumes the structure of singularities in the CGSMM 
is much richer than in the SMM, and they match in the limit $V \rightarrow \infty$ only.
To see this let us first make the inverse Laplace transform:
\begin{eqnarray}\label{eight}
{\cal Z}(V,T,\mu) &= &
\int\limits_{\chi - i\infty}^{\chi + i\infty}
\frac{ d \lambda}{2 \pi i} ~ \hat{\cal Z}(\lambda,  T, \mu)~ e^{\textstyle   \lambda \, V } =
%
\int\limits_{-\infty}^{+\infty} \hspace*{-0.1cm} d \xi
\int\limits_{-\infty}^{+\infty} \hspace*{-0.1cm}  \frac{d \eta}{{2 \pi}}  
\hspace*{-0.1cm} \int\limits_{\chi - i\infty}^{\chi + i\infty}
\hspace*{-0.1cm} \frac{ d \lambda}{2 \pi i}~ 
\frac{\textstyle e^{ \lambda \, V - i \eta \xi } }{{\textstyle \lambda - i\eta ~-~{\cal F}(\xi,\lambda - i\eta)}}~= 
\nonumber \\
&& \int\limits_{-\infty}^{+\infty} \hspace*{-0.1cm} d \xi
\int\limits_{-\infty}^{+\infty} \hspace*{-0.1cm}  \frac{d \eta}{{2 \pi}}
\,{\textstyle e^{  i \eta (V - \xi)  } } \hspace*{-0.1cm} \sum_{\{\lambda _n\}}
e^{\textstyle  \lambda _n\, V } 
{\textstyle 
\left[1 - \frac{\partial {\cal F}(\xi,\lambda _n)}{\partial \lambda _n} \right]^{-1} } \,,
\end{eqnarray}
where the contour  $\lambda$-integral is reduced to the sum over the residues of all singular points
$ \lambda = \lambda _n + i \eta$ with $n = 1, 2,..$, since this  contour in the complex $\lambda$-plane  obeys the
inequality $\chi > \max(Re \{  \lambda _n \})$.  
Now both remaining integrations in (\ref{eight}) can be done, and the GCP becomes 
\begin{equation}\label{nine}
{\cal Z}(V,T,\mu)~ = \sum_{\{\lambda _n\}}
e^{\textstyle  \lambda _n\, V }
{\textstyle \left[1 - \frac{\partial {\cal F}(V,\lambda _n)}{\partial \lambda _n} \right]^{-1} } \,,
\end{equation}
i.e. the double integral in (\ref{eight}) simply  reduces to the substitution   $\xi \rightarrow V$ in
the sum over singularities. 
In  \cite{Bugaev:05a} this  remarkable result was  
formulated as a theorem, which now is generalized to more complicated forms of the liquid phase pressure.


The simple poles in (\ref{eight}) are defined by the  equation 
\begin{equation}\label{ten}
\lambda _n~ = ~{\cal F}(V,\lambda _n)\,.
\end{equation}
In contrast to the usual SMM \cite{Bugaev:00} the singularities  $ \lambda _n $ 
are (i) 
 are volume dependent functions, if $K(V)$ is not constant,
and (ii) they can have a non-zero imaginary part, but 
in this case there  exist  pairs of complex conjugate roots of (\ref{ten}) because the GCP is real.

Introducing the real $R_n$ and imaginary $I_n$ parts of  $\lambda _n = R_n + i I_n$,
we can rewrite  Eq. (\ref{ten})
as a system of coupled transcendental equations 
\begin{eqnarray}\label{eleven}
&&\hspace*{-0.2cm} R_n = ~ \sum\limits_{k=1}^{K(V) } \tilde\phi_k (T)
~{\textstyle e^{\frac{Re( \nu_n)\,k}{T} } } \cos(I_n b k)\,,
\\
\label{twelve}
&&\hspace*{-0.2cm} I_n = - \sum\limits_{k=1}^{K(V) } \tilde\phi_k (T)
%
%
~{\textstyle e^{\frac{Re( \nu_n)\,k}{T} } } \sin(I_n b k)\,,
\end{eqnarray}
where we have introduced the 
set of  the effective chemical potentials  $\nu_n  \equiv  \nu(\lambda_n ) $ with $ \nu(\lambda) = p_l(T,\mu) b  - \lambda b\,T$, and 
the reduced distributions $\tilde\phi_1 (T) = \left(\frac{m T }{2 \pi} \right)^{\frac{3}{2} }  z_1 \exp((\mu -  p_l(T,\mu) b)/T)$ and 
$\tilde\phi_{k > 1} (T) = \left(\frac{m T }{2 \pi} \right)^{\frac{3}{2} }  k^{-\tau}\, \exp(-\sigma (T)~ k^{2/3}/T)$ for convenience.

Consider the real root $(R_0 > 0, I_0 = 0)$, first. 
Similarly to the SMM, 
for $I_n = I_0 = 0$ the real root $R_0$  of  the GSMM exists for any $T$ and $\mu$.
Comparison  of  $R_0$ from (\ref{eleven}) with the expression for vapor pressure of the analytical SMM solution 
\cite{Bugaev:00}
indicates  that $T R_0$ is  a constrained grand canonical pressure of the mixture of ideal gases with the chemical potential $\nu_0$. 
 As usual,  for  finite volumes the total mechanical pressure \cite{Hill} differs from   $T R_0$.
Equation (\ref{twelve}) shows that for $I_{n>0} \neq 0$ the inequality $\cos(I_n b k) \le 1$ cannot   simultaneously
become an equality for all $k$-values. Then from Eq. (\ref{eleven})  
one obtains ($n>0$)
\begin{equation}\label{thirteen}
R_n < \sum\limits_{k=1}^{K(V) } \tilde\phi_k (T)
~{\textstyle e^{\frac{Re(\nu_n)\, k}{T} } } \quad \Rightarrow \quad R_n < R_0\,, 
\end{equation}
where the second inequality (\ref{thirteen}) immediately follows from the first one.
 In other words, the gas singularity is always the rightmost one.
This fact 
plays a decisive role in the thermodynamic limit
$V \rightarrow \infty$.

The interpretation of the complex roots $\lambda _{n>0}$  seems to be  less straightforward and, hence, in this case  we follow  the line of arguments suggested  in  Ref.  \cite{Bugaev:05a}. 
According to Eq. (\ref{nine}),   the  GCP is a superposition of  the
states of different  free energies $- \lambda _n V T$.  
Strictly speaking,  $- \lambda _n V T$  has  a meaning of  the change of free energy, but
we  will use the  traditional  term for it.
For $n>0$ the free energies  are complex. 
Therefore,  
 $-\lambda _{n>0} T$ is   the density of free energy.  The real part  of the free energy density,  $- R_n T$, defines the significance of the state's  contribution to the partition:  due to (\ref{thirteen}) the  largest contribution  always comes from the gaseous state and has the lowest value of the  real part  of free energy density. As usual,  the states which do not correspond to the lowest  value of the (real part of)  free energy, i. e.  $- R_{n>0} T$, are thermodynamically metastable. 
For  infinite   volume they should not contribute unless they are infinitesimally close to  $- R_{0} T$,  but for finite volumes their contribution to the GCP may be important.

As one can see from (\ref{eleven}) and (\ref{twelve}), the states of  different  free energies have  
different values of the effective chemical potential $\nu_n$, which is not the case for
infinite volume \cite{Bugaev:00},
where there  exists a single value for the effective chemical potential. 
Thus,  for finite $V$
the  states which contribute to the GCP (\ref{nine}) are not in a true chemical equilibrium.

The meaning of the imaginary part of the free energy density  becomes  clear from 
(\ref{eleven}) and (\ref{twelve}): as it is  seen from (\ref{eleven})   the imaginary part $I_{n>0}$
effectively changes the number of degrees of freedom of  each $k$-nucleon fragment ($k \le K(V)$)
contribution to  the free energy  density  $- R_{n>0} T$.  It is clear, that the change 
of the effective number of degrees of freedom can occur virtually only and, if 
$\lambda _{n>0}$ state is accompanied  by 
some kind of  equilibration process. 
Both of these statements become clear,
 if we recall that  
the statistical operator in statistical mechanics and the  quantum mechanical evolution operator 
are related by the Wick rotation \cite{Feynmann}. In other words, the inverse temperature can be
considered as an  imaginary time.  
Therefore, depending on the sign,  the quantity  $ I_n b T \equiv \tau_n^{-1}$  that  appears 
 in the trigonometric  functions      
of  the  equations (\ref{eleven}) and (\ref{twelve}) in front of the imaginary time $1/T$ 
can be regarded  as the inverse decay/formation time $\tau_n$ of the metastable state which corresponds to the  pole $\lambda _{n>0}$ (for more details see next sections and  \cite{Bugaev:05a}).
 
Such an  interpretation of  $\tau_n$ naturally explains the thermodynamic  metastability 
of all states except the gaseous one:
the metastable states can exist  in the system only virtually 
because of their finite decay/formation  time,
whereas the gaseous state is stable because it has an infinite decay/formation time.

%
%
%
%
\begin{figure}[ht]
\includegraphics[width=8.6cm]{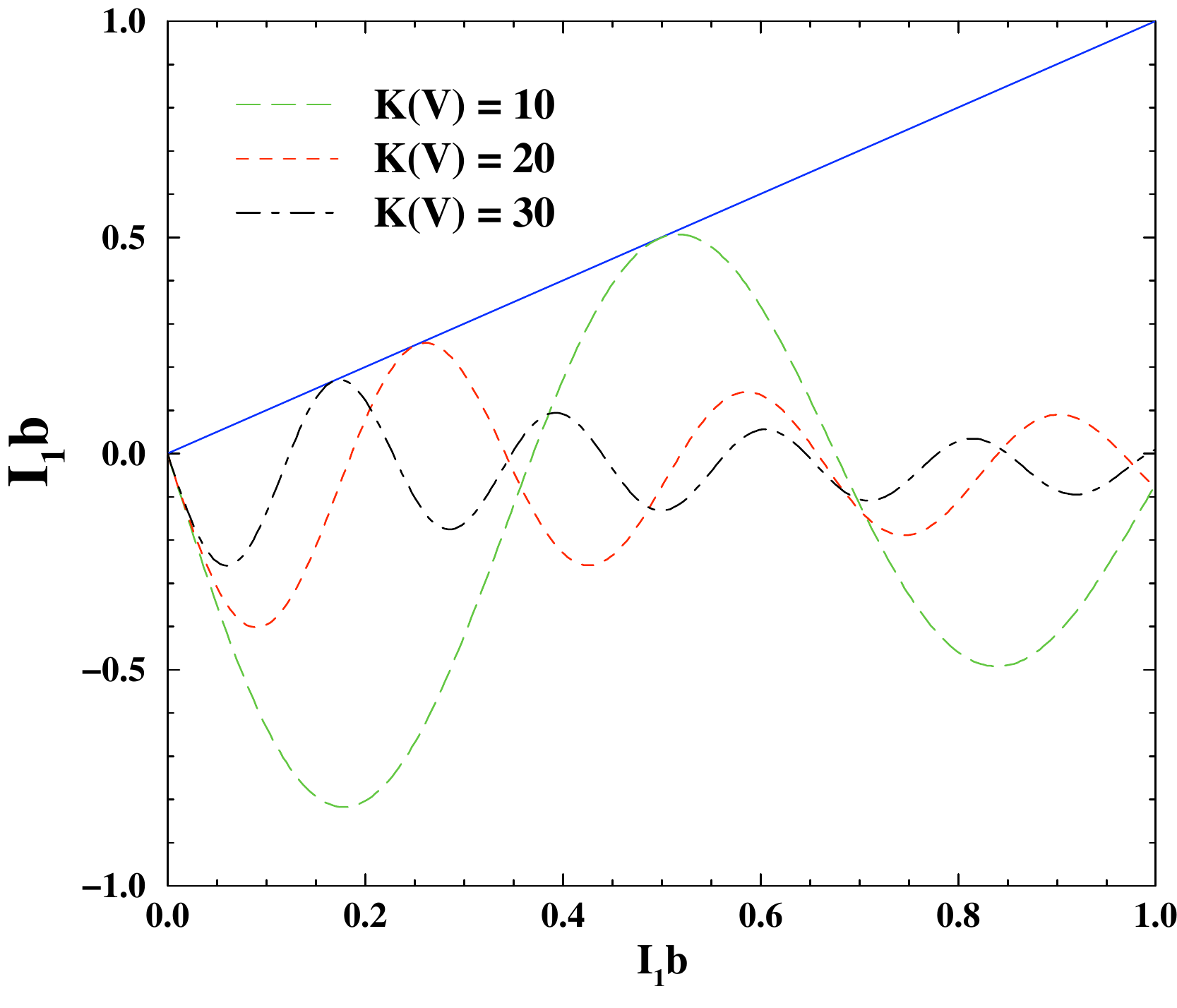}
\caption{A graphical solution of Eq. (\ref{twelve}) for $T = 10$ MeV and $\tau = 1.825$ for the typical SMM parameterization of the surface tension coefficient by Eq. (\ref{one}).
Note, however, that qualitatively the same picture remains valid for any parameterization 
of the surface tension coefficient. 
The l.h.s. (straight line) and  r.h.s. of Eq. (\ref{twelve}) (all dashed curves) are shown
as the function of
dimensionless parameter $I_1\,b$ for the three values of the largest fragment size $K(V)$.
The intersection point at $(0;\,0)$ corresponds to a real root of Eq. (\ref{ten}).
Each tangent point with the straight line generates  two complex  roots of (\ref{ten}).
}
  \label{fig1}
\end{figure}

\section{No Phase Transition Case}


It is instructive to treat the effective chemical potential $\nu (\lambda)$ as an independent variable
instead of $\mu$. In contrast to the infinite $V$, where  the upper   limit  $\nu \le 0$ defines the liquid phase singularity of the isobaric partition and  gives the pressure of a liquid phase
$p_l(T,\mu) = T R_0 |_{V \rightarrow \infty} $, 
for finite  volumes and finite $K(V)$ the effective  chemical potential can
be complex (with either sign for its real part)  and its value defines the number and position of the imaginary roots $\{\lambda _{n>0} \}$  in the complex plane.
Positive  and negative values of the effective chemical potential  for finite systems  were  considered 
within the Fisher droplet model \cite{Elliott:01}, but, to our  best knowledge,  its complex values have   been  discussed for the first time in \cite{Bugaev:05a}.
From the definition of the effective chemical potential $\nu(\lambda)$ it is evident that  its complex
values for finite systems  exist  only  because of the excluded volume interaction, which is 
not taken into account in the  Fisher droplet model \cite{Fisher:67}.

As it is seen from  Fig.~1, the r.h.s. of Eq. (\ref{twelve})  
is the amplitude and frequency modulated sine-like  
function of dimensionless parameter $I_n\,b$. 
Therefore, depending on $T$ and $Re(\nu)$ values, there may exist  no 
complex roots $\{\lambda _{n>0}\}$, a finite number of them, or an infinite number of them. 
In Fig.~1 we showed a special case which corresponds to  exactly three 
roots of Eq. (\ref{ten}) for each value of $K(V)$: the real root ($I_0 = 0$) and two complex conjugate
roots ($\pm I_1$). 
Since 
the r.h.s. of (\ref{twelve}) is monotonously increasing function
of  $Re(\nu)$,  
it is possible to map the $T-Re(\nu)$ plane into
regions of a fixed number of roots of Eq. (\ref{ten}). 
For fixed $T$-value  each curve in \mbox{Fig. 2} divides the $T-Re(\nu)$ plane
into three parts:  for $Re(\nu)$-values below the curve $Re(\nu_1 (T))$ there  is only one real root (gaseous phase), 
for points on  the curve $Re(\nu) = Re(\nu_1 (T))$ there exist      
three roots, and above the curve $Re(\nu_1 (T))$  there are five or more roots of Eq. (\ref{ten}).
Although \mbox{Fig. 2} corresponds to the usual SMM parameterization of the surface tension coefficient,
the  picture is qualitatively the same for general parameterization of $\sigma$ whereas  its modifications are discussed below. 

For constant values of  $K(V) \equiv K$  the number of terms in the r.h.s. of (\ref{twelve}) does 
not depend on the volume and, consequently, in thermodynamic limit $V \rightarrow \infty$   only the 
rightmost  simple pole in the complex $\lambda$-plane survives out of a finite number of simple poles.
According to the  inequality (\ref{thirteen}), the real root $\lambda_0$ is  the  rightmost singularity of isobaric partition (\ref{six}).
However,  there is a  possibility  that the real parts  of other  roots $\lambda_{n>0} $ become infinitesimally close to $R_0$, when there is an infinite number of terms which contribute to the GCP (\ref{nine}).

%
%
 \begin{figure}[ht]
\includegraphics[width=8.6cm,height=6.0cm]{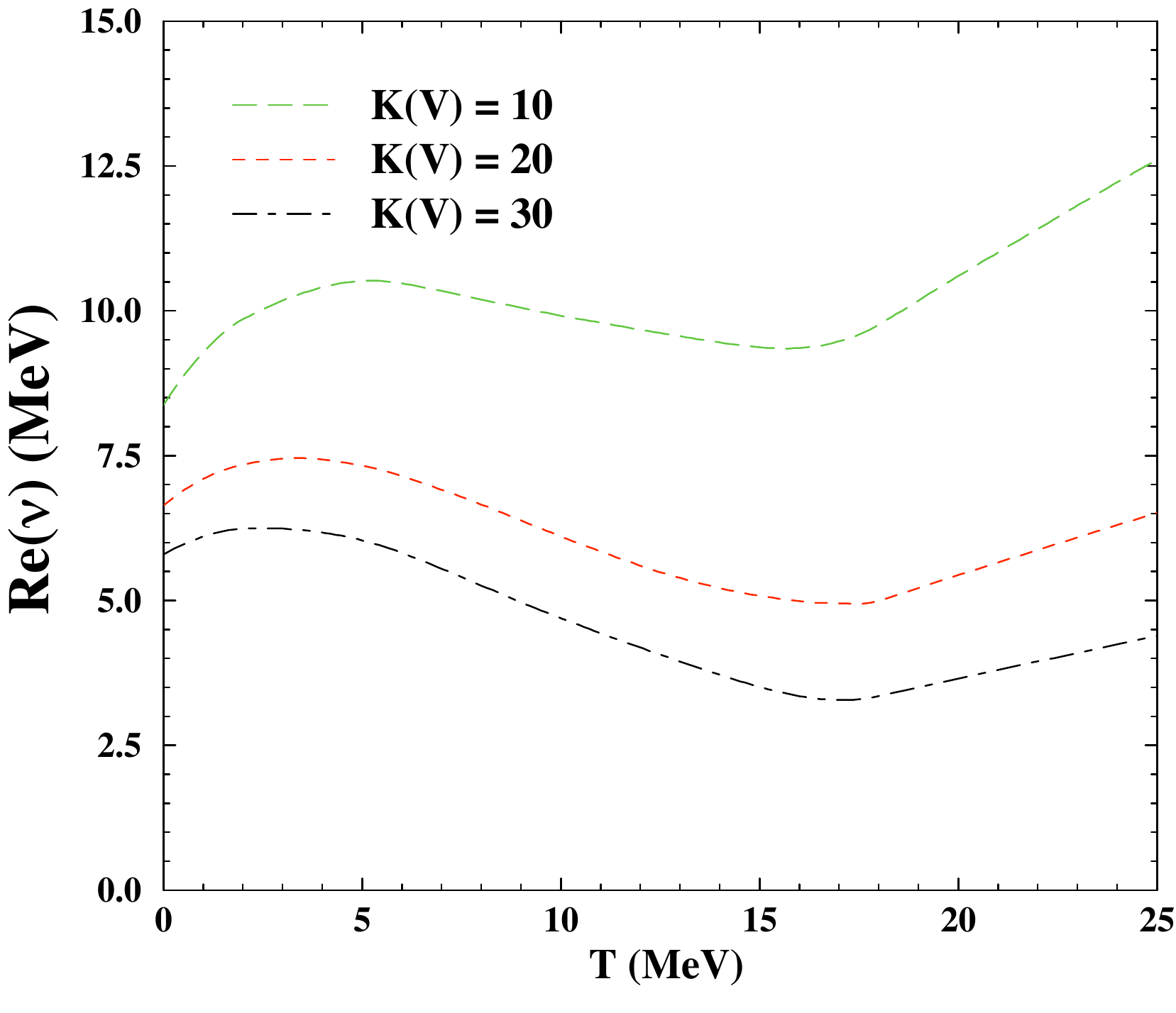}
  \caption{Each curve separates  the $T-Re(\nu_n)$ region of one real root of Eq. (\ref{ten})
(below the curve), three complex roots (at the curve) and five and more roots (above the curve)
for three values of $K(V)$ and the same parameters as in Fig. 1.
}
  \label{fig2}
\end{figure}

Let us show  now that even for an infinite number of simple poles in (\ref{nine})
only the real root $\lambda_0$  survives in the limit $V \rightarrow \infty$.
For this purpose consider  the limit $Re(\nu_n)  \gg T $.
In this limit   the distance between  the imaginary parts of the nearest roots 
remains finite even for infinite volume.  Indeed,  for $Re(\nu_n)  \gg T   $
the leading contribution to the r.h.s. of (\ref{twelve}) corresponds to the harmonic with $k = K$,
and, consequently,  an exponentially large amplitude of this term
can be only  compensated by  a vanishing value of  $\sin\left( I_n \, b K  \right)$,  
i.e.   $I_n \, b K  =  \pi n + \delta_n$
with $|\delta_n| \ll \pi$ (hereafter we will analyze only 
the branch $I_n > 0$), 
and, therefore, the corresponding decay/formation time $\tau_n  \approx K [ \pi n T ]^{-1}$ 
is volume independent.

Keeping the leading term on the r.h.s. of  (\ref{twelve}) and solving for $\delta_n$, one finds
\begin{eqnarray}\label{Mfourteen}
\hspace*{-0.2cm}
I_n & \approx & \frac{\pi\, n + \delta_n}{b \, K(v)} \,, \\
\label{Mfifteen}
%
\delta_n &\approx  &  \frac{ (-1)^{n+1}  \pi n }{ K b ~ \tilde\phi_K (T)  }~{
\textstyle e^{- \frac{Re(\nu_n )\,K}{T} } } \,,    \\
\label{Msixteen}
R_n & \approx & (-1)^{n}  \tilde\phi_K (T)  ~{\textstyle e^{\frac{Re(\nu_n )\,K}{T} } }  \,, 
\end{eqnarray}
where in the last step we used Eq. (\ref{eleven}) and condition $|\delta_n| \ll \pi$.
Since for $V \rightarrow \infty$ all  negative values of $R_n$ cannot contribute to the 
GCP (\ref{nine}), it is sufficient to analyze even values of $n$ which, according to 
(\ref{Msixteen}),  generate  $R_n > 0$.

Since the inequality  (\ref{thirteen}) can not be broken,   a single possibility,
when $\lambda_{n>0}$ pole can contribute to the partition (\ref{nine}),  corresponds to 
the case 
$ R_n \rightarrow R_0 - 0^+$ 
for  some finite $n$.   
Assuming this, we find  $Re (\nu (\lambda_n)) \rightarrow Re(\nu (\lambda_0))$
for the same value of $\mu$. 

Substituting these results into equation (\ref{eleven}), one gets
\begin{equation}\label{Mseventeen}
\hspace*{-0.2cm}R_n \approx  \sum\limits_{k=1}^{K } \tilde\phi_k (T)
~{\textstyle e^{\frac{Re(\nu (\lambda_0) )\,k}{T} } } \cos\left[ \frac{ \pi n k}{K} \right]  \ll R_0\,.
\end{equation}
The inequality (\ref{Mseventeen})  follows  from the equation for $R_0$ and the fact that, even for
equal leading terms   in the sums above (with $k =  K$ and even  $n$),  the difference between $R_0$  and $R_n$ is  large due to  the next to leading term $k = K - 1$, which is  proportional to 
$e^{\frac{Re(\nu (\lambda_0) )\,(K-1)}{T} } \gg 1$.  
Thus, we arrive at  a  contradiction with our assumption $R_0 - R_n \rightarrow 0^+$, 
and, consequently,  it cannot be true.  Therefore,
for large volumes  the  real root $\lambda_0$  always gives
the main contribution to  the GCP (\ref{nine}), and  this is the only root that survives 
in the limit $V \rightarrow \infty$.
Thus,  we showed that  the model with the fixed  size  of the largest fragment has no phase transition because there is a single singularity of the isobaric partition (\ref{six}), which 
exists in thermodynamic limit. However, for the finite systems we can also define the analog of the metastable  mixed phase which corresponds to a finite number of complex conjugate solutions $\lambda_{n>0}$. 
Clearly, that in thermodynamic limit  the contribution of these metastable states into all physical quantities disappears. 
The equation $Re(\nu) = Re(\nu_1(T))$ defines the boundary between the finite volume analogs of the  gaseous and mixed  phases in $T-Re(\nu)$ and   $T-\mu$ planes.

\section{Finite Volume Analogs of Phases}
 
If $K(V)$ monotonically grows with the volume,  the situation is different. 
In this case for  positive value of $Re(\nu)  \gg T$ 
the leading exponent in the r.h.s. of (\ref{twelve})  
also corresponds to a largest fragment, i.e. to $k  = K(V)$. 
Therefore,  we can apply 
the same arguments  which were used above  for  the case $K(V) = K =  const$
and derive similarly  equations  (\ref{Mfourteen})--(\ref{Msixteen}) for  $I_n$ and $R_n$. 
From the relation  $I_n  \approx \frac{\pi n}{ b\, K( V) }$ it follows that,
when $V$ increases, the number of simple poles in (\ref{eight}) also increases 
and
the  imaginary part of 
the closest to the real $\lambda$-axis  poles becomes very small,
 i.e $I_n   \rightarrow 0$ for  $n \ll K(V)$,
 and, consequently, the associated   decay/formation time 
$\tau_n  \approx K(V)  [ \pi n T ]^{-1}$ grows with the volume of the system.
Due to the fact that  $ I_n  \rightarrow 0$, 
the inequality (\ref{Mseventeen}) cannot be  
established for the poles with $n \ll K(V)$. 
Therefore, in  contrast to the previous case, for large $K(V)$ the  simple poles
with $n \ll K(V)$ will be infinitesimally close to the real axis of the complex $\lambda$-plane.

In this case from Eq.  (\ref{Msixteen}) one obtains 
\begin{eqnarray}
T R_n & \approx &  p_l(T,\mu)  -  \frac{ T}{ K(V) b} 
 \ln \left|  \frac{ R_n }{  \tilde\phi_K (T)     }  \right|  \nonumber \\
 & \approx &  p_l(T,\mu) - \frac{\sigma}{[K(V)]^\frac{1}{3} b} - T \left[ \frac{\ln |
 \left(\frac{2 \pi}{ m T } \right)^{\frac{3}{2}} R_n  | + \tau \ln K(V)}{K(V) b} \right]
  \label{Meighteen}
\end{eqnarray}
for  $ Re( \nu ) \gg T $ and $K(V) \gg 1$.   Thus, 
from Eq. (\ref{Meighteen}) one can clearly see   that
for an  infinite volume an   infinite number of simple poles moves toward 
the real $\lambda$-axis to the vicinity of liquid phase singularity $\lambda_l = p_l(T,\mu)/T $ 
of the isobaric partition
\cite{Bugaev:00} and
generates  an essential singularity of function ${\cal F}(V, p_l/T)$ in (\ref{seven}) 
{\it irrespective to the  sign of the liquid phase pressure $p_l(T,\mu)$.}
As  we showed above, the states with $Re( \nu ) \gg T$  become  stable because they acquire   infinitely large 
decay/formation time $\tau_n$ in the limit $V \rightarrow \infty$.  
Therefore, these states should be identified 
as a liquid phase for finite  volumes as well. 
Such a conclusion can be easily understood, if we recall that
the  partial pressure $T R_n$  of  (\ref{Meighteen})
corresponds to a single  fragment of the largest possible size. 
Moreover, as one can see from the leading terms on the r.h.s. of (\ref{Meighteen}) 
the partial pressure $T R_n$  contains both the liquid phase and the  surface contributions
for a spherical fragment of the mean radius $[K(V) b]^\frac{1}{3}$.  
In fact, the above results remain  valid under a weaker  condition  $ Re( \nu ) K(V) \gg T $ 
 since such an inequality allows one to  establish  the approximation (\ref{Msixteen}).

Now it is clear 
that each curve in Fig.~2  is   the  finite volume analog of the phase boundary $T-\nu$ for a given value of $K(V)$:  below the phase boundary there exists a gaseous phase, but at and above each curve there
are  states which can be identified with a finite volume analog of the mixed phase, and,
finally, if  $Re( \nu )/T \rightarrow \infty$  there exists a liquid phase.
Again as in the previous section the equation $Re(\nu) = Re(\nu_1(T))$ defines the boundary between the finite volume analogs of the  gaseous and mixed  phases in $T-Re(\nu)$ and   $T-\mu$ planes. Clearly, for finite $V$ the solution of this equation $\mu_c (T, V)$
depends on $T$ and $V$. 

Although the calculations depicted in Fig.~2 were made for $\sigma (T) \ge 0$ and for 
finite values of the effective chemical potential $Re( \nu )$ the shown results can be qualitatively explained using Eq. (\ref{Meighteen}) in   the limit $Re( \nu ) K (V) /T \rightarrow \infty$.  Indeed, from  (\ref{Meighteen}) one finds
\begin{eqnarray}
Re( \nu_n ) \approx  \frac{\sigma}{[K(V)]^\frac{1}{3} } + T \left[ \frac{\ln |
 \left(\frac{2 \pi}{ m T } \right)^{\frac{3}{2}} R_n  | + \tau \ln K(V)}{K(V) } \right] \, ,
  \label{EqXXII}
\end{eqnarray}
from which one obtains $Re( \nu_n ) > 0$ for finite $K(V)$ values  (compare to Fig.~2), if  $|\left(\frac{2 \pi}{ m T } \right)^{\frac{3}{2}} R_n  | \gg 1 $ and $\sigma \ge 0$.  
If, however, the surface tension coefficient gets negative, i.e.  $\sigma < 0$ for $T > T_c$, then for sufficiently large values of $K(V)$ one can find that  $Re( \nu_n ) <  0$, i.e. in this case  the finite volume analog of phase boundary  can demonstrate another behavior than 
that one shown in Fig.~2. For $T = T_c$ the surface tension coefficient vanishes and from 
(\ref{EqXXII}) we get
\begin{eqnarray}
Re( \nu_n ) \biggl|_{T=T_c} \approx \frac{ T_c }{K(V) } \left[  \ln \left|
 \left(\frac{2 \pi}{ m T_c } \right)^{\frac{3}{2}} \frac{ p_l(T_c,\mu )}{ T_c}  \right| + \tau \ln K(V)  
\right] 
\, , \label{EqXXIII}
\end{eqnarray}
where in the last step of derivation we replaced $R_n$ by the leading term from the r.h.s. of (\ref{Meighteen}). This result shows that (i) at $T=T_c$   the deviation of the partial  pressure $T R_n$ from the liquid phase pressure  decreases faster  as  function of $K(V)$ than for other temperatures, but  at the same time  
(ii) for large  $K(V)$ this  deviation still  decreases   slower than   the imaginary part $I_n$.  For a quantitative  example  let us choose $\mu$ in such a way that the first term on the r.h.s. of (\ref{EqXXIII}) disappears, i.e. for   $Re( \nu_n ) \bigl|_{T=T_c} \approx \frac{ T_c  \tau}{K(V) }  \ln K(V)  $. Then for $\tau = 1.825$ one finds  
$Re( \nu_n ) \bigl|_{T=T_c}  \approx 7.56$ MeV for $K(V) =10$,
$Re( \nu_n ) \bigl|_{T=T_c}  \approx 4.92$ MeV for $K(V) =20$,
$Re( \nu_n ) \bigl|_{T=T_c}  \approx 3.72$ MeV for $K(V) =30$,
 and $Re( \nu_n ) \bigl|_{T=T_c}  \approx 1.5$ MeV for $K(V) =100$. 
From Fig.~2 one can see that, although  our  estimate of $Re( \nu_n ) \bigl|_{T=T_c}$ for $K(V) =10$ is about 2.5 MeV below its value found  numerically, the corresponding  estimates for  $K(V) =20$ and $K(V) =30$ obtained from 
(\ref{EqXXIII}) are in a very good  agreement with the results of the numerical evaluation.

When  there is no phase transition, i.e. $K(V) = K = const$,  the structure of simple poles is
similar, but, first,  the line which separates the gaseous states from the metastable states does not change with the system volume, and, second, as shown above, the metastable states will never become stable.  Therefore, a systematic study of the 
volume dependence  of free energy (or pressure for very large  $V$)  along with the formation and  decay times may  be  of a crucial importance for  experimental studies of 
the nuclear liquid gas phase transition.

The above results demonstrate that, in contrast  to Hill's expectations \cite{Hill}, the finite volume analog of the mixed phase does not  consist  just of  two pure phases. 
The mixed phase for finite volumes
consists of  a stable  gaseous phase and  the set of  metastable states which have different   free energy values. 
Moreover, the difference between the free energies of these states is  not  the surface-like, as Hill 
assumed in his treatment \cite{Hill}, but the  volume-like as we have seen.  Furthermore, 
according to Eqs. (\ref{eleven}) and (\ref{twelve}),  each of these states 
consists of the same fragments, but with different weights. 
As was  shown above for the case $ Re(\nu) \gg T$,  
some fragments 
that  belong to 
the states, in which  the largest fragment is  dominant,
may, in principle, have negative weights (effective number of degrees of freedom) in 
the expression  for  $R_{n>0}$  (\ref{eleven}).
This can be understood easily because  higher concentrations of large fragments can be achieved 
 at the  expense of the  smaller fragments and  is reflected in  the corresponding change 
of the real part of  the free energy $- R_{n>0} V T$. 
Therefore, the  actual  structure of the mixed phase at finite volumes is  more complicated
than  was expected in earlier works.  

Here it is necessary to add a few remarks  about the description of the deconfinement PT 
on the basis of statistical  models \cite{QGBSTM1,QGBSTM2,FWM:08,Aleksey:11} that were  solved recently in thermodynamic limit. The finite volume solution  of the models \cite{QGBSTM1,QGBSTM2,FWM:08,Aleksey:11} can be straightforwardly found using the LFT 
developed in \cite{Bugaev:07a,Bugaev:05a}. Recently, however, the  importance  of finite width of heavy/large quark gluon plasma bags  was realized \cite{FWM:08, Reggeons1, Reggeons2}. Both the theoretical estimates \cite{Reggeons1, Reggeons2} and the  analysis of the asymptotic  Regge trajectories of  non-strange mesons \cite{Bugaev:11} indicate 
that  the  width of the quark gluon  bag  of the volume $V $  with the mass $M$ being  heavier than $M_0 \approx 2.5$ GeV  is $\Gamma = \gamma (T) \sqrt{ \frac{V}{V_0} }$ (here $V \ge V_0 = 1$ fm$^3$). Since even at $T=0$ the  value  $\gamma (T=0) \approx 400$ MeV is large,  we conclude that the short life time $t_{life} (V) = 1/\Gamma(V)$ of such bags can, in principle,  modify our conclusions about the metastable states $\lambda_{n>0}$.  This is so because in the SMM and the  GSMM  the fragments are  implicitly assumed to be  stable and, hence, their   life-time is set  to be  infinite while   in reality there is the region of stability of nuclear fragments outside of which the nucleus  life-time is extremely short.  A similar situation is with large/heavy quark gluon bags although 
the instability of nuclei is due to the Coulomb interaction whereas the short life-time of the bags should be  attributed to such a property of strongly interacting matter as the color confinement \cite{FWM:08,String:10,String:11}. 

Now it is clear that, if the individual life-time of the largest bag $t_{life} (V)$ in the volume $V$ is larger than the collective decay/formation time of the state $\lambda_{n>0}$ which for 
quark gluon bags is $\widetilde\tau_n \approx \frac{V}{V_0 \pi n T}$ \cite{Bugaev:07a}, then the largest bag can be considered as a stable one during the course of  collective decay/formation process, i.e. $t_{life} (V) \ge  \widetilde\tau_n $. Otherwise the process of  collective decay or formation of the state $\lambda_{n>0}$ cannot  ever be completed because the largest bag ceases its existence 
much earlier.  Note that under such conditions  one can hardly expect an existence of thermal equilibrium in the system of shortly living particles. 

Therefore, the inequality $t_{life} (V) \ge  \widetilde\tau_n $ sets some constraints on 
the  applicability range  of the above analysis to the short-living bags 
\begin{eqnarray}
\left[ \frac{V}{V_0} \right]^\frac{3}{2} & \le & \frac{\pi \, T}{\gamma(T)} \, n
\, . \label{EqXXIV}
\end{eqnarray}
This inequality can be used as an estimate for the volume of  largest bags in a finite system, if $n$ and $T$ are known. For instance, if $\gamma(T) = \gamma(T=0) = const$, then one obtains $\frac{V}{V_0} \le \left[ \frac{\pi \, T}{\gamma(T=0)} \, n \right]^\frac{2}{3}$. 
Alternatively,  Eq. (\ref{EqXXIV}) can be used to estimate the range of temperatures 
for given $n$ since $V/V_0 \ge 1$. Using the last inequality one can rewrite (\ref{EqXXIV}) as $n \ge \frac{\gamma(T)}{ \pi\, T}$. Substituting into the last result  the following expression for $\gamma(T) = 
\sqrt{2 V_0 \,  a\, T T_H^2 ( T^2 + TT_H + T_H^2)}$ that  was predicted in \cite{Reggeons1},
where $T_H \approx 160$ MeV is the Hagedorn temperature and $a= \frac{37}{90} \pi^2$ for SU(3) color group with two quark  flavors,  we find 
$n \ge  1$ for $T \ge  500$ MeV and $n \ge 2$ for $T \ge 1120$ MeV. These estimates show that at LHC energies we can expect an existence of metastable states  which can be described within the developed  approach. 


\section{Conclusions}

In this work we generalized the SMM to GSMM and included into it   the nuclear liquid phase pressure of the most general form. This allowed us to get rid of the absolute  incompressibility of the nuclear liquid. Also  in this model  we introduced very  general form of the surface tension coefficient which enabled us to repair another pitfall of the simplified  SMM related to the absence of critical endpoint in this model for $\tau >2$. 
The LFT method was applied to the constrained GSMM
and this model was solved analytically  for  finite volumes.

 It is shown that  for finite volumes the GCP function 
can be identically rewritten in terms of the simple poles of the isobaric partition (\ref{six}).
The real pole $\lambda_0$  exists always and  the quantity  $T \lambda_0$ is the GCP  pressure of the gaseous
phase.  The complex roots $\lambda_{n>0}$ appear as pairs of complex conjugate solutions
of equation (\ref{ten}).   As we discussed, their  most straightforward interpretation  is as follows: 
$- T Re (\lambda_{n>0}) $ has a meaning of 
the free energy density, whereas $ b T Im (\lambda_{n>0} )$, depending on  sign,  gives
the  inverse  decay/formation time  of such a state.
The gaseous state is always stable  because its decay/formation time is infinite and because
its free energy is minimal. 
The complex poles describe the metastable states for $Re( \lambda_{n>0} ) \ge  0 $ and mechanically
unstable states for $Re( \lambda_{n>0} ) < 0 $.

We studied the volume dependence of the simple poles and found a dramatic difference
in their  behavior  in case with PT  and  without it.
For the former
one the found  representation  allows one  to define the finite volume analogs of phases  unambiguously 
and to establish the finite volume analog of the $T-\mu$  phase diagram (see Fig. 2). 
At finite volumes the gaseous phase exists, if there is  a single simple pole, the mixed phase corresponds to three and more simple poles, 
whereas the liquid is represented by an infinite amount of simple poles at highest possible 
particle density (or $\mu \rightarrow \infty$).

As we showed, for given $T$ and $\mu$ the states of the mixed phase which have different  
$Re( \lambda_{n} )$ are
not in a true chemical equilibrium for finite volumes. This feature cannot be obtained within 
the Fisher droplet model due to lack
of the hard core repulsion between fragments.  This fact also  demonstrates clearly  that, 
in contrast to Hill's expectations \cite{Hill}, the mixed phase is  not just  a composition of two 
states which are the pure phases. 
As we showed, the mixed phase is a superposition of three and more collective states, 
and each of them 
is characterized by its own  value of $\lambda_n$.
Because of that  the difference between the  free energies  of these states is not the  surface-like,
as Hill argued \cite{Hill}, but the  volume-like.

A detailed analysis of the isobaric partition singularities in the $T - Re (\nu)$ plane allowed us to define the finite volume analogs of  phases and  study  the behavior of these singularities  in the limit 
$V \rightarrow \infty$.  Such an analysis  opens a possibility  to rigorously   study  the nuclear liquid-gas PT and the deconfinement 
PT   directly from the finite volume partition. This may help to extract 
the phase diagram of strongly interacting matter  
from the experiments on finite systems (nuclei)  with more confidence.
The conditions of the model applicability to the description of the short-living quark gluon plasma bags are also discussed. 

\bigskip

{\bf  Acknowledgments.}
The  authors appreciate the fruitful  discussions with D. B. Blaschke, J. B. Elliott,  O. V.  Teryaev and D. N. Voskresensky.   
K.A.B., A.I.I. and G.M.Z. acknowledge the partial support 
of  the Program ``Fundamental Properties of Physical Systems 
under Extreme Conditions'' launched by  the Section of Physics and Astronomy  of
the National Academy of Sciences of Ukraine. 
The work of E.G.N. and A.S.S.  was supported in part by the Russian
Foundation for Basic Research, Grant No. 11-02-01538-a.


\end{document}